\documentclass[12pt]{iopart}
\usepackage{iopams} 
\usepackage{graphicx}

\begin{document}

\title[Gravitational polarization]{Gravitational polarization and \\the
phenomenology of MOND}

\author{Luc Blanchet}

\address{
${\mathcal{G}}{\mathbb{R}}\varepsilon{\mathbb{C}}{\mathcal{O}}$,
Institut d'Astrophysique de Paris, UMR 7095 du CNRS, Universit\'e
Pierre \& Marie Curie, $98^\mathrm{bis}$ boulevard Arago, 75014 Paris,
France} \ead{blanchet@iap.fr}

\begin{abstract}
The modified Newtonian dynamics (MOND) has been proposed as an alternative to
the dark matter paradigm; the philosophy behind is that there is no dark
matter and we witness a violation of the Newtonian law of dynamics. In this
article, we interpret differently the phenomenology sustaining MOND, as
resulting from an effect of ``gravitational polarization'', of some cosmic
fluid made of dipole moments, aligned in the gravitational field, and
representing a new form of dark matter. We invoke an internal force, of
non-gravitational origin, in order to hold together the microscopic
constituents of the dipole. The dipolar particles are weakly influenced by the
distribution of ordinary matter; they are accelerated not by the gravitational
field, but by its gradient, or tidal gravitational field.
\end{abstract}

\maketitle

\section{Introduction}\label{sec1}
The observed discrepancy between the dynamical mass and the luminous mass of
bounded astrophysical systems is generally attributed to the existence of an
invisible form of matter, coined as the missing mass or dark matter. The
nature of the dark matter particles is unknown but extensions of the standard
model of particle physics provide a number of candidates~\cite{BHS05}. The
dark matter triggers the formation of large-scale structures by gravitational
collapse and predicts the scale-dependence of density
fluctuations. Simulations suggest some universal dark matter density profile
around ordinary matter distributions~\cite{NFW96}. An important characteristic
of dark matter, required by the necessity of clustering matter on small
scales, is that it should be cold or non-relativistic at the epoch of the
galaxy formation. However, the dark matter hypothesis has some
difficulties~\cite{McG98} at explaining naturally the flat rotation curves of
galaxies, one of the most persuasive evidence for the existence of dark
matter, and the Tully-Fisher empirical relation~\cite{TF} between the observed
luminosity and the asymptotic rotation velocity of spiral galaxies.

On the other hand, the modified Newtonian dynamics (MOND) has been proposed by
Milgrom~\cite{Milg1,Milg2,Milg3} as an alternative to the dark matter
paradigm. It imputes the mass discrepancy not to the presence of some
additional non-baryonic matter, but to failure of the Newtonian law of gravity
in an appropriate regime --- a drastic change of paradigm. MOND involves a
single parameter $a_0$ being a constant acceleration scale, which delineates
the specific MOND regime, corresponding to accelerations much smaller than
$a_0$, from the Newtonian regime, for which the accelerations are much
larger. Several relativistic extensions of MOND, assuming the existence of
extra fields associated with gravity, besides the spin-$2$ metric field of
general relativity, have been proposed~\cite{BekM84,BekS94}. Such extensions
have culminated in the scalar-vector-tensor theory of Bekenstein and
Sanders~\cite{Sand97,Bek04,Sand05}.

MOND has been very successful at fitting the flat rotation curves of galaxies,
and at recovering naturally the Tully-Fisher relation
(see~\cite{Milgrev,SMcG,Bekrev,McG,Sandrev} for reviews). Intriguingly, the
numerical value of $a_0$ that fits the data is close to the Hubble scale,
$a_0\approx c\,H_0$. MOND may have also some observational problems: There are
some counter-examples of galaxies where MOND does not seem to account for the
observed kinematics~\cite{Gent}, and, most importantly, the mass discrepancy
at the scale of clusters of galaxies is not entirely explained by
MOND~\cite{Gerbal,Clowe}.

In this article and the next one~\cite{B06dipole} (hereafter paper~II), we
take the view that MOND does not represent a violation of the fundamental law
of gravity, but, rather, provides us with an important hint on the (probably
unorthodox) nature of the elusive dark matter. More precisely, we interpret
the phenomenology behind MOND as resulting from an effect of gravitational
polarization, of some cosmic fluid made of dipole moments, and representing a
new form of dark matter. The dipole moments get aligned in the gravitational
field produced by ordinary masses, thereby enhancing the magnitude of the
field and yielding MOND. Such effect is the gravitational analogue of the
usual electrostatic effect, of polarization of a dielectic medium in an
applied electric field~\cite{jackson}. 

In the present paper we imagine, as a model for the dipole, a doublet of
particles, one having a positive gravitational mass, the other having a
negative and opposite gravitational mass, and with both particles being
endowed with positive inertial masses. The gravitational behavior involving
masses of this type is governed by a negative Coulomb law --- like masses
attract and unlike masses repel~\cite{Bondi57}. As a result the dipole moment
cannot be stable: Even if we neglect the repulsive gravitational force between
the two particles, they will accelerate apart from each other in an exterior
gravitational field produced by ordinary matter. We shall therefore invoke an
internal force, of non-gravitational origin, between the two particles
constituting the dipole, to bound them in a gravitational field. The MOND
acceleration scale $a_0$ will appear to be related to the properties of this
internal ``microscopic'' force at short distances. We find that the motion of
the dipolar particles violates the equivalence principle, and is driven by the
\textit{tidal} gravitational field of ordinary matter, rather than the
gravitational field itself. In this sense the dark matter is only weakly
influenced by the distribution of ordinary matter.

Summarizing, in our approach the dark matter is described by a
``digravitational'' medium, which is subject to polarization in a
gravitational field, and is otherwise essentially static (an ``ether''). An
alternative interpretation of this dark matter is by the gravitational
analogue of a plasma in electromagnetism, \textit{i.e.} composed of positive
and negative gravitational masses, and oscillating at the natural plasma
frequency. In a gravitational field the mean position of the masses is
displaced from equilibrium and the plasma acquires a dipolar polarization. The
observational predictions of the present (non-relativistic) model are the same
as for MOND.

In paper~II we shall propose a relativistic model of dipolar particles, based
on an action principle in general relativity. This model will be consistent
with the equivalence principle, and as a result the dynamics of dipolar
particles, even in the non-relativistic limit, will be different from that of
the present quasi-Newtonian model. The present paper and paper~II provide two
distinct models, both of them suggest a close connection between the
phenomenology of MOND and some form of gravitationally polarized dipolar dark
matter.

Section~\ref{sec2} investigates the formal analogy between the MOND equation
and the electrostatics of non-linear media. Sections~\ref{sec3} and~\ref{sec4}
introduce a microscopic quasi-Newtonian description of the gravitational
dipole moment. In Sec.~\ref{sec5} we derive the expression of the
non-gravitational internal force in the MOND regime. In Sec.~\ref{sec6} we
present an alternative though equivalent formulation of the dipolar medium in
terms of a polarized gravitational plasma.

\section{Analogy with electrostatics}\label{sec2}
The MOND equation, in the variant derivable from a non-relativistic
Lagrangian~\cite{BekM84}, takes the form of the modified Poisson equation
\begin{equation}\label{MONDeq}
\partial_i\!\left[ \mu\!\left(\frac{g}{a_0}\right)g^i\right] = -4\pi \,G
\,\rho\,,
\end{equation}
where $\rho$ denotes the density of ordinary matter, $g^i=\partial_i U$ is the
gravitational field in the non-relativistic limit (so that $a^i=g^i$ is the
acceleration of ordinary matter), and $U$ is the gravitational
potential.\footnote{Spatial indices $i,j$ are raised and lowered using the
Euclidean metric $\delta_{ij}$; the summation convention is used throughout.}
In Eq.~(\ref{MONDeq}) the Milgrom function $\mu$ depends on the ratio $g/a_0$,
where $g=\vert g^i\vert$ is the norm of the gravitational field, and $a_0$ is
the constant acceleration scale. The MOND regime corresponds to the limit of
weak gravity, much below the scale $a_0$, \textit{i.e.} $g\ll a_0$; in this
limiting regime we have $\mu(g/a_0)\approx
g/a_0$~\cite{Milg1,Milg2,Milg3}. When $g\gg a_0$, the function $\mu(g/a_0)$
asymptotes to one, and we recover the usual Newtonian law. Sometimes we shall
consider the formal ``Newtonian'' limit $g\rightarrow\infty$; however we
always assume that the gravitational field is non-relativistic. Various forms
of the function $\mu$ have been proposed (\textit{cf.}~\cite{FB05}), but most
of them appear to be rather \textit{ad hoc}.

Taking the MOND equation~(\ref{MONDeq}) at face, we notice a striking
analogy with the usual equation of electrostatics describing the
electric field inside a dielectric medium, namely $\partial_i
D^i=\rho_\mathrm{e}$, where the electric induction $D^i$ is
proportional to the electric field $E^i$ (at least for not too large
electric fields): $D^i = \mu_\mathrm{e} \,\varepsilon_0 \,E^i$. Here
$\mu_\mathrm{e}=1+\chi_\mathrm{e}$ is the dielectric coefficient, and
$\chi_\mathrm{e}$ denotes the electric susceptibility of the
dielectric medium, which depends on the detailed microscopic
properties of the medium (see \textit{e.g.}~\cite{jackson}). In
non-linear media the susceptibility is a function of the norm of the
electric field, $\chi_\mathrm{e}(E)$ with $E=\vert E^i\vert$. The
electric polarization is proportional to the electric field, and is
given by $\Pi_\mathrm{e}^i = \chi_\mathrm{e} \,\varepsilon_0
\,E^i$. The density of electric charge due to the polarization is
$\rho^\mathrm{pol}_\mathrm{e}=-\partial_i\Pi_\mathrm{e}^i$.
Generically we have $\chi_\mathrm{e}>0$, which corresponds to
\textit{screening} of electric charges by the polarization charges,
and reduction of the electric field inside the dielectric.

In keeping with this analogy, let us interpret the MOND function $\mu$
entering Eq.~(\ref{MONDeq}) as a ``digravitational'' coefficient, and
write
\begin{equation}\label{muchi}
\mu = 1 + \chi\,,
\end{equation}
where $\chi$ would be a coefficient of ``\textit{gravitational
susceptibility}'', parametrizing the relation between some
``\textit{gravitational polarization}'', say $\Pi^i$, and the
gravitational field:
\begin{equation}\label{Pigi}
\Pi^i = - \frac{\chi}{4\pi \,G} \,g^i\,.
\end{equation}
Since as we have seen the MOND function~(\ref{muchi}) depends on the
magnitude of the gravitational field, $\mu(g/a_0)$, the same is true
of the so defined gravitational susceptibility, $\chi(g/a_0)$, in
close analogy with the electrostatics of non-linear media. Hence we
expect that $\chi$ should characterize the response of some non-linear
digravitational medium to an applied gravitational field. The mass
density associated with the polarization would then be given by the
same formula as in electrostatics,
\begin{equation}\label{rhopol}
\rho_\mathrm{pol}=-\partial_i \Pi^i\,.
\end{equation}
With those notations Eq.~(\ref{MONDeq}) can be rewritten as
\begin{equation}\label{MONDpol}
\Delta U = -4\pi \,G\,(\rho+\rho_\mathrm{pol})\,.
\end{equation}
In such rewriting of MOND, we see that the Newtonian law of gravity is
not violated, but, rather, we are postulating the existence of a new
form of matter, to be called dark matter, and which contributes in the
normal way to the right-hand-side (RHS) of the Poisson
equation~(\ref{MONDpol}). The dark matter consists of polarization
masses with volume density $\rho_\mathrm{pol}$ given
by~(\ref{rhopol}).

\section{Sign of the susceptibility coefficient}\label{sec3}
For the moment we have restated the MOND equation in the form
(\ref{rhopol})--(\ref{MONDpol}) and proposed a formal
interpretation. To check this interpretation, let us view the
digravitational medium as consisting of individual dipole moments
$\pi^i$ with number density $n$, so that the polarization vector reads
\begin{equation}\label{Pipi}
\Pi^i = n\,\pi^i\,.
\end{equation}
We suppose that the dipoles are made of a doublet of sub-particles, one with
positive mass $+m$ and one with negative mass $-m$, the masses which we are
referring here are the \textit{gravitational masses} of these particles,
\textit{i.e.} the gravitational analogue of the electric charges,
$m_\mathrm{g}=\pm m$. Clearly the exotic nature of this dark matter shows up
here, when we suggest the notion of negative gravitational masses. If the two
masses $\pm m$ are separated by the spatial vector $d^i$, pointing in the
direction of the positive mass, the dipole moment is
\begin{equation}\label{pid}
\pi^i = m\,d^i\,.
\end{equation}
Let us further suppose that the two sub-particles are endowed with
\textit{inertial masses} which are \textit{positive}, and given by
$m_\mathrm{i}=m$. The dipole moment thus consists of an ordinary
particle, say $(m_\mathrm{i},m_\mathrm{g})=(m,m)$, associated with an
exotic particle, $(m_\mathrm{i},m_\mathrm{g})=(m,-m)$. 

The ordinary particle will always be attracted by an external mass
distribution made of ordinary matter; however, the other particle
$(m_\mathrm{i},m_\mathrm{g})=(m,-m)$ will always be \textit{repelled}
by the same external mass. In addition the two sub-particles will
repel each other. We see therefore that the gravitational dipole is
unstable, and we shall invoke a non-gravitational force to sustain
it. We also expect that the external gravitational field will exert a
torque on the dipole moment in such a way that its orientation has the
positive mass $+m$ oriented in the direction of the external mass, and
the negative mass $-m$ oriented in the opposite direction. We thus
find that $\pi^i$ and $\Pi^i$ should both point towards the external
mass, \textit{i.e.} be oriented in the same direction as the external
gravitational field $g^i$. From Eq.~(\ref{Pigi}) we therefore conclude
that the susceptibility coefficient $\chi$, in the gravitational case,
must be \textit{negative}:
\begin{equation}\label{chisign}
\chi < 0\,.
\end{equation}
This corresponds to an ``\textit{anti-screening}'' of the ordinary mass by the
polarization masses, and \textit{enhancement} of the gravitational field in
the presence of the digravitational medium. This simply results from the fact
that, in contrast to electrostatics, alike gravitational charges or masses
always attract. The result~(\ref{chisign}) is nicely compatible with the
prediction of MOND; indeed we have $0\leq\mu<1$ in a straightforward
interpolation between the MOND and Newtonian regimes, hence
$-1\leq\chi<0$. The stronger gravitational field predicted by MOND may thus be
naturally interpreted by a process of anti-screening by polarization masses.

\section{Microscopic model for the dipole}\label{sec4}
To give more substance to the model, suppose that some interaction $F^i$
between the two constituents of the dipole is at work, and let the dipole
moment be embedded into the gravitational field $g^i=\partial_i U$. The
equations of motion of the sub-particles $(m_\mathrm{i},m_\mathrm{g})=(m,m)$
and $(m,-m)$, having positions $x_{+}^i$ and $x_{-}^i$ respectively, are
\begin{eqnarray}
m\,\frac{d^2 x_{+}^i}{d t^2} &=& m\,g^i(x_{+})-F^i(x_{+}-x_{-})\,,\label{xplus}\\
m\,\frac{d^2 x_{-}^i}{d t^2} &=& -m\,g^i(x_{-})+F^i(x_{+}-x_{-})\,.\label{xmoins}
\end{eqnarray}
The internal force $F^i$ is proportional to the relative separation vector
$d^i = x_{+}^i-x_{-}^i$, namely
\begin{equation}\label{Fi}
F^i = F\,\frac{d^i}{d}\,.
\end{equation}
The norm of $F^i$ is a function of the separation distance, $F=F(d)$ where $d
= \vert x_{+}^i-x_{-}^i\vert$, and is expected to also depend on the magnitude
of the gravitational field, $g=\vert g^i\vert$. The force $F^i$ is assumed to
be \textit{attractive}, $F>0$ (hence it is non-gravitational). This force is
indispensable if we are looking for configurations in which the constituents
of the dipole remain at constant distance from each other. For simplicity, we
suppose that the gravitational force between the two sub-particles,
\textit{i.e.}  $F_\mathrm{g}=-G \,m^2/d^2$, which is repulsive, is negligible
or included into the definition of $F$.

We introduce next the centre of \textit{inertial} masses, $x^i =
(x_{+}^i+x_{-}^i)/2$, and transform the system of
equations~(\ref{xplus})--(\ref{xmoins}) into an \textit{equation of motion}
for the ``dipolar particle'',
\begin{equation}\label{EOM}
2 m\,\frac{d^2 x^i}{d t^2} = \pi^j \partial_{ij} U +
\mathcal{O}\left(d^2\right)\,,
\end{equation}
and an \textit{evolution equation} for the dipole moment,
\begin{equation}\label{evoleq}
\frac{d^2 \pi^i}{d t^2} = 2 m\,g^i - 2 F^i +
\mathcal{O}\left(d^2\right)\,.
\end{equation}
In both Eqs.~(\ref{EOM}) and~(\ref{evoleq}) we neglect terms of the order of
the square of the separation distance $d$, assuming that $d\ll\vert
x^i\vert$. In the RHS of~(\ref{EOM})--(\ref{evoleq}), $g^i$ and $\partial_{ij}
U$ are evaluated at the position of the center-of-mass $x^i$. The torque
produced by the forces acting on the two sub-particles in the RHS
of~(\ref{xplus})--(\ref{xmoins}) is given by
\begin{equation}\label{torque}
C^i=\varepsilon^{ijk}\pi^j g^k+ \mathcal{O}(d^2)\,.
\end{equation}
This torque will tend to align the dipole moment with (and in the same
direction of) the gravitational field.

The prominent feature of the equation of motion~(\ref{EOM}) is the violation
of the equivalence principle by the dipolar particle, as a result of the fact
that the particle's inertial mass is $2m$ while its gravitational mass is
zero. Note that Eq.~(\ref{EOM}) has a structure different from the equation we
obtain in the non-relativistic limit of the relativistic model of paper~II ---
indeed, contrarily to the present quasi-Newtonian model, the relativistic
model is consistent with the equivalence principle (see paper~II for
discussion). While the inertial and passive/active gravitational masses of the
dipolar particle are given here by $M_\mathrm{i}=2m$ and
$M_\mathrm{p}=M_\mathrm{a}=0$, the relativistic model of paper~II will have
$M_\mathrm{i}=M_\mathrm{p}=2m$ and $M_\mathrm{a}=\mathcal{O}(c^{-2})$ in the
non-relativistic limit $c\rightarrow\infty$. This represents a fundamental
difference between the two models. In addition, of course, since the present
model is Newtonian it does not \textit{a priori} allow (in contrast to
paper~II) to answer questions related to cosmology or the motion of
relativistic particles. However, despite these differences, we shall recover
in paper~II the main characteristics of the mechanism of gravitational
polarization.

From Eq.~(\ref{EOM}) the dipolar particle is expected to accelerate slowly in
a given gravitational field, as compared to an ordinary particle. More
precisely, we find that the particle is not directly subject to the
gravitational field, but, rather, to its gradient, namely the \textit{tidal
gravitational field} $\partial_{ij} U$. In the potential $U\sim 1/R$
(respectively the MOND analogue $U\sim \ln R$), the acceleration is typically
of the order of $1/R^3$ (resp. $1/R^2$).\footnote{This type of coupling to the
tidal gravitational field is well known; for instance it corresponds to the
non-relativistic limit of the coupling between the spin and the Riemann
curvature tensor, for particles with spin moving on an arbitrary
background~\cite{Papa51spin,BI80}.}  The observational consequence is that the
dark matter consisting of a fluid of dipole moments is necessarily cold, and
even ``colder'' than ordinary non-relativistic matter. This property may be
consistent with the observation of galactic structures. Thus, the dipolar dark
matter appears as a medium whose dynamics is weakly influenced by the
distribution of ordinary galaxies.

On the other hand, the evolution equation~(\ref{evoleq}) shows that a
situation of equilibrium, where the distance $d$ between the pair of particles
remains constant, is possible. The equilibrium is realized when the internal
force $F^i$ exactly compensates for the gravitational force,\footnote{In
Sec.~\ref{sec6} we shall be more precise about what is meant by equilibrium
condition; see Eq.~(\ref{Fiav}).}
\begin{equation}\label{Feq}
F^i = m\,g^i + \mathcal{O}\left(d^2\right)\,.
\end{equation}
Note that here as everywhere else, $g^i$ represents the \textit{total}
gravitational field, sum of the contributions due to the ordinary masses and
the polarization masses. Because $F^i$ is proportional to $d^i$,
Eq.~(\ref{Fi}), we see that when the equilibrium holds, the dipole moment is
in the direction of the gravitational field, $\pi^i=m\,d^i\propto g^i$. The
polarization vector $\Pi^i=n\,\pi^i$ is aligned with the gravitational field
and the medium is polarized. Hence, the equilibrium condition~(\ref{Feq})
provides a mechanism for verifying the crucial equation~(\ref{Pigi}).

Let us now get information on the susceptibility coefficient $\chi$ as a
function of $g=\vert g^i\vert$. From Eq.~(\ref{Pigi}) and the relation
$\Pi=n\,m\,d$ [see~(\ref{Pipi})--(\ref{pid})], we get
\begin{equation}\label{suscept}
\chi = - 4\pi \,G\,m\,n\,\frac{d}{g}\,.
\end{equation}
%
The condition~(\ref{Feq}) implies that $d$ is necessarily a function of $g$,
say $d=d(g)$, obtained by inversion of the relation $F(d)=m\,g$ or, rather,
since as we have seen the force should also depend on $g$, of the relation
$F(d,g)=m\,g$. Thus, $\chi$ is a certain function of $g$, depending on the
properties of the internal force $F^i$, and we are able, in principle, to
relate the MOND function $\mu=1+\chi$ to the internal structure of the dipolar
particles. We obtain
\begin{equation}\label{mug}
\mu\!\left(\frac{g}{a_0}\right) = 1 - 4\pi \,G\,m\,n\,\frac{d(g)}{g}\,.
\end{equation}
Notice that such function is expected to be a complicated function of $g$,
because it is made of the inverse of $F(d,g)=m\,g$, and especially because it
depends on the spatial distribution of the dipole moments, characterized by
their number density $n$. The distribution of $n$ is determined by the
gravitational field \textit{via} the equation of motion~(\ref{EOM}), together
with the Eulerian continuity equation $\partial_t n + \partial_i ( n v^i)=0$,
where $v^i = d x^i/dt$. However, as we have seen, the motion of the dipolar
particle is sensitive in first approximation only to the tidal gravitational
field. Thus, a reasonable approximation is probably to consider that the
velocity field $v^i$ remains small, hence the number density $n$ is nearly
constant and uniform. In the following we shall neglect the tidal
gravitational fields, and shall treat $n$ as a constant.

\section{Internal force law}\label{sec5}
In the limit $g\rightarrow 0$, we enter the deep MOND regime --- a non-linear
regime characterized by $\mu = g/a_0 + \mathcal{O}(g^2)$.\footnote{For
simplicity we assume a power-law expansion when $g\rightarrow 0$.} Comparing
with Eq.~(\ref{mug}), we deduce that the dipole separation $d$ should behave
in terms of the gravitational field in the MOND regime like
\begin{equation}\label{dg}
d = \frac{g}{4\pi \,G\,m\,n}\left[1 - \frac{g}{a_0} +
\mathcal{O}\left(g^2\right)\right]\,.
\end{equation}
Thus, in first approximation, $d$ is found to be proportional to the
gravitational field (recall that $n$ is assumed to be constant); this means,
using $F=m\,g$, that the force must be dominantly proportional to $d$. More
precisely, we find that the force law $F(d)$ that is necessary to account for
the MOND phenomenology is
\begin{equation}\label{Fd}
F(d) = 4\pi \,G\,m^2\,n\,d \left[1 + \frac{4\pi \,G\,m\,n}{a_0} \,d +
\mathcal{O}\left(d^2\right)\right]\,.
\end{equation}
Interestingly, this force becomes weaker when the particles constituting the
dipole moment get closer. As we see from~(\ref{Fd}), the MOND acceleration
scale $a_0$ happens to parametrize, in this model, the expansion of the
internal force at short distances, \textit{i.e.} $d\rightarrow 0$ (in the
regime where $g\rightarrow 0$).

Equation~(\ref{Fd}) represents the value of the force at equilibrium,
\textit{i.e.} when~(\ref{Feq}) is satisfied, and it depends on the number $n$
of particles. However the internal force $F^i$ itself --- not necessarily at
equilibrium --- is defined by the equations of
motion~(\ref{xplus})--(\ref{xmoins}) for a single dipole moment without
reference to $n$, and in this sense is \textit{intrinsic} to the dipole
moment. Nevertheless it seems unusual that the equilibrium force~(\ref{Fd})
should depend on the surrounding density $n$ of the medium. In Sec.~\ref{sec6}
we shall show that this force is actually the one corresponding to an harmonic
oscillator describing the oscillations of a ``gravitational plasma'' at its
natural plasma frequency. The dependence on $n$ of Eq.~(\ref{Fd}) will then
appear to be that involved into the usual expression~\cite{jackson} of the
plasma frequency, as given by Eq.~(\ref{omega}) below.

In addition, we want to recover the usual Newtonian gravity when $g\gg a_0$,
and we see from Eq.~(\ref{mug}) that it suffices that $d(g)/g\rightarrow 0$
when $g\rightarrow\infty$. There is a large number of possibilities; many
force laws $F(d)$ do it in practice. For instance we can adopt $d\propto
g^{1-\epsilon}$ with any $\epsilon>0$, which corresponds to the power force
law $F(d)\propto d^\alpha$ with $\alpha=1/(1-\epsilon)$, and we see that any
powers $\alpha$, except those with $0\leq\alpha\leq 1$, are possible. In the
discussion below we shall choose $\epsilon>1$ in order to ensure that
$d\rightarrow 0$ in the Newtonian regime. The particular case $\epsilon=3/2$
gives back the $1/d^{2}$ type force law. In this case we can adopt for the
susceptibility function
\begin{equation}\label{chiN}
\chi = - \left(\frac{a_0}{g}\right)^{3/2}\left[1 +
\mathcal{O}\left(\frac{1}{g}\right)\right]\,.
\end{equation}
This corresponds to the dipolar separation
\begin{equation}\label{dginf}
d = \frac{a_0}{4\pi \,G\,m\,n}\left(\frac{a_0}{g}\right)^{1/2}\left[1 +
\mathcal{O}\left(\frac{1}{g}\right)\right]\,,
\end{equation}
and internal force
\begin{equation}\label{Fdinf}
F = m\,a_0\left(\frac{a_0}{4\pi\,G\,m\,n\,d}\right)^2 \Bigl[1 +
\mathcal{O}\left(d^2\right)\Bigr]\,.
\end{equation}
Notice that because the expression~(\ref{Fdinf}) is positive, it represents a
Coulombian force, \textit{i.e.} attractive between unlike masses.

Since $d$ tends to zero in both the Newtonian and MOND regimes, we see that
the function $d\rightarrow F(d)$ is actually two-valued. We have already
noticed that $F$ depends not only on $d$ but also on $g$; the expression of
the force~(\ref{Fd}) is valid in the MOND regime where $g\rightarrow 0$, while
Eq.~(\ref{Fdinf}) holds in the Newtonian regime where $g\rightarrow\infty$. An
alternative (but rather \textit{ad hoc}) choice, encompassing both types of
behavior, is provided by the susceptibility function $\chi=-e^{-g/a_0}$. In
this case we have
\begin{equation}\label{dginf0}
d = \frac{g}{4\pi \,G\,m\,n}\,e^{-g/a_0}\,,
\end{equation}
and the force law $F(d)$ at equilibrium is obtained by substituting
$g=F/m$ in the RHS and looking for the two-valued inverse.

%
%
%
\begin{figure}[h]
\centerline{\includegraphics[width=8cm]{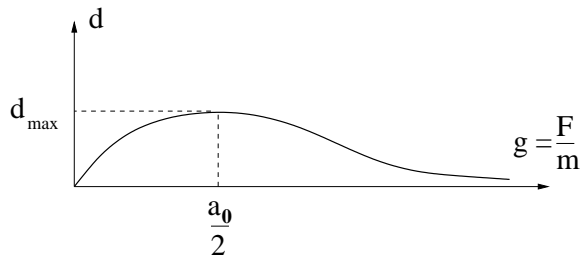}}
\vspace{-0.7cm}\caption{The dipolar separation distance $d$ as a function of
$g$. The equilibrium condition $F=m\,g$ is satisfied. Such graph is valid when
the number density $n$ of dipole moments is constant. More generally, for non
constant $n$, the physically meaningful analogue of $d$ would be the
polarization $\Pi=n\,m\,d$ (\textit{cf.} Fig.~4 in paper~II).}
\label{fig1}\end{figure}
The generic form of the distance function $d(g)$ is illustrated in
Fig.~\ref{fig1}. From the figure we comment on the physical picture one might
have in mind. In the absence of gravitational fields, \textit{i.e.} in the
absence of ordinary matter, the dipole moments do not exist (at least
classically) since their separation $d$ is zero.  Indeed, when $g=0$ we have
$d=0$ by Eq.~(\ref{dg}). The dipolar ether does not produce any noticeable
effect. Suppose that some external mass, made of ordinary matter, is steadily
approached. The dipole moments start feeling a weak gravitational field and
they become active. According to Eq.~(\ref{dg}) and Fig.~\ref{fig1}, the
dipoles open up and get aligned in the gravitational field in order to
maintain the equilibrium~(\ref{Feq}). The medium is polarized, we are in the
MOND regime, and the gravitational field is dominated by the contribution of
the polarization masses. Further approaching the ordinary mass, the dipolar
separation $d$ eventually reaches a maximal value. From~(\ref{dg}) we find,
approximately,
\begin{equation}\label{dmax}
d_\mathrm{\max}\approx \frac{a_0}{16\pi\,G\,m\,n}\,,
\end{equation}
which is reached for $g\approx a_0/2$. If at that point we continue to
increase the gravitational field (putting nearer the external mass), $d$ will
begin to decrease and the dipole moments will close up. Finally, for strong
gravitational fields, the dipole moments become inactive again (indeed $d=0$
when $g\rightarrow\infty$). The gravitational field is dominated by the
contribution of the ordinary matter, and we recover the Newtonian regime.
\section{Gravitational plasma}\label{sec6}
The medium of dipolar dark matter described above can also be interpreted, in
an alternative formulation, as a polarizable ``gravitational plasma''
consisting of the two species of particles $(m_\mathrm{i},m_\mathrm{g})=(m,\pm
m)$. Such formulation will essentially be equivalent to the previous one, but
is simpler and more appealing physically.

Suppose that the two types of particles $(m,\pm m)$ are in equal numbers so
that the plasma is globally neutral. There is an equilibrium point where the
plasma is \textit{locally} neutral, so the number \textit{densities} of the
two particle species are equal, at each point, to some common constant and
uniform value $n$. As it stands the equilibrium is unstable because the
gravitational force between unlike masses is repulsive. To ensure a stable
equilibrium we must postulate like in Eqs.~(\ref{xplus})--(\ref{xmoins}) some
restoring non-gravitational force $F^i$, acting between the masses
$m_\mathrm{g}=\pm m$, and superseding the gravitational force. We introduce
some associated internal field $f^i$ such that
\begin{equation}\label{Fifi}
F^i = - m f^i\,,
\end{equation}
and assert that this field obeys a Gauss law in the non-relativistic limit,
\begin{equation}\label{divf}
\partial_i f^i = -\frac{4\pi\,G\,m}{\chi}\,\bigl(n_{+}-n_{-}\bigr)\,.
\end{equation}
The number densities of the particles species $(m,\pm m)$ are denoted by
$n_{\pm}$. We have introduced here the susceptibility coefficient $\chi$ to
represent a dimensionless coupling constant characterizing the internal
interaction. Notice that since $\chi$ is negative, Eq.~(\ref{chisign}), the
force law~(\ref{divf}) is attractive between unlike particles and repulsive
between like ones. Furthermore, supposing that the plasma is bathed by an
external gravitational field $g^i$, constant and uniform over some region of
consideration,\footnote{We adopt the frame associated with the plasma's
equilibrium configuration. The gravitational field $g^i$ is defined in that
frame.} we expect that the coupling constant should reflect the presence of
this gravitational field, and we assume a dependence on its norm $g$, namely
$\chi=\chi(g)$.

Let $x^i_{+}$ and $x^i_{-}$ (in short $x^i_{\pm}$) be the displacement vectors
of the masses from the equilibrium position characterized by the density
$n$. The particles are accelerated by the internal field $f^i$ and by the
applied external gravitational field $g^i$. The equations of motion have
already been given in Eqs.~(\ref{xplus})--(\ref{xmoins}) and read now
\begin{equation}\label{eom}
m\,\frac{d^2 x^i_{\pm}}{d t^2} = \pm \,m\,\bigl(f^i + g^i\bigr)\,.
\end{equation}
Consider a small departure from equilibrium, corresponding to small
displacement vectors $x^i_{\pm}$. The density perturbations are given by
$n_{\pm} = n\,(1-\partial_i x^i_{\pm})$ to first order in the displacements
$x^i_{\pm}$. Using Eq.~(\ref{divf}) we readily integrate for $f^i$ and inject
the solution into~(\ref{eom}). In this way we find that $x^i_{+}+x^i_{-}=0$
(the centre of inertial masses is at rest --- neglecting tidal fields),
together with the following harmonic oscillator for $\pi^i = m\,d^i$ where
$d^i = x^i_{+}-x^i_{-}$,
\begin{equation}\label{harmosc}
\frac{d^2 \pi^i}{d t^2} + \omega^2\,\pi^i = 2m\,g^i\,.
\end{equation}
Actually this computation is the classic derivation of the plasma
frequency~\cite{jackson} which is found for the case at hands to be
\begin{equation}\label{omega}
\omega = \sqrt{-\frac{8\pi\,G\,m\,n}{\chi}}\,.
\end{equation}
The frequency depends on the density $n$ of the plasma at equilibrium and on
the strength of the internal interaction which is encoded into the coupling
constant $\chi$. The solution we get for the internal field is
\begin{equation}\label{exprFi}
f^i = -\frac{\omega^2}{2m}\,\pi^i\quad\Longrightarrow\quad F^i =
\frac{\omega^2}{2}\,\pi^i\,,
\end{equation}
so we can check that Eq.~(\ref{harmosc}) is equivalent to our previous
equation~(\ref{evoleq}) (recall that here $g^i$ is considered to be constant
and uniform). The particles oscillate around some non-zero mean position that
is determined by the ambiant gravitational field as
\begin{equation}\label{piav}
\langle \pi^i\rangle = \frac{2m}{\omega^2}\,g^i\,.
\end{equation}
The mean value of the force is
\begin{equation}\label{Fiav}
\langle F^i\rangle = m\,g^i\,,
\end{equation}
and coincides with the equilibrium condition postulated in Eq.~(\ref{Feq}).

This result~(\ref{piav}) is classical but can easily be recovered from
standard quantization of the harmonic oscillator~(\ref{harmosc}). The spectrum
of ``plasmons'' is discrete ($k\in\mathbb{N}$) with energy levels shifted by
the gravitational field~\cite{cohen},
\begin{equation}\label{Ekp}
E_k = \left(k+\frac{3}{2}\right)\,\hbar\,\omega - \frac{m\,g^2}{2\omega^2}\,.
\end{equation}
The eigenstates are $\vert\psi_{p_1p_2p_3}\rangle=
\vert\phi_{p_1}\rangle\vert\phi_{p_2}\rangle\vert\phi_{p_3}\rangle$ where
$k=p_1+p_2+p_3$; the eigenvalues have degeneracy $(k+1)(k+2)/2$. The
one-dimensional eigenstate function $\phi_{p_i}(\pi^i)=\langle
\pi^i\vert\phi_{p_i}\rangle$ is of the form $\varphi_{p_i}(\pi^i-\frac{2
m}{\omega^2}g^i)$ where ($H_p$ being the Hermite polynomial)
\begin{equation}\label{phip}
\varphi_{p}(\rho) =
\frac{1}{2^{\frac{p}{2}}\sqrt{p!}}\left(\frac{\omega}{\pi\,\hbar\,m}\right)^{1/4}
H_p(\hbox{$\sqrt{\frac{\omega}{\hbar\,m}}$}\,\rho)\,e^{-\frac{\omega}{2\hbar\,m}\rho^2}\,.
\end{equation}
The expectation value $\langle \pi^i\rangle= \langle\psi_{p_1p_2p_3}\vert
\pi^i\vert\psi_{p_1p_2p_3}\rangle$ does not vanish due to the presence of the
gravitational field; it can be computed using
$\int_{-\infty}^{+\infty}d\pi\,\pi\,\vert\varphi_p\bigl(\pi-\frac{2
m}{\omega^2}g\bigr)\vert^2 = \frac{2 m}{\omega^2}g$ and the result agrees
with~(\ref{piav}).

We analyze the effect of the mean dipole moment vector~(\ref{piav}) on the
equation for the gravitational field $g^i$ (supposed now to be generated by
external sources). More precisely we argue that the mean value is really a
quantum expectation value as we have just proved, so that by adopting a
``semi-classical'' approach this expectation value $\langle \pi^i\rangle$
should have a direct effect as the source for the \textit{classical}
gravitational field $g^i$. The polarization readily follows from the
expression of the plasma frequency~(\ref{omega}) as
\begin{equation}\label{respolar}
\Pi^i = n\,\langle \pi^i\rangle = - \frac{\chi}{4\pi \,G} \,g^i\,.
\end{equation}
so that we recover exactly Eq.~(\ref{Pigi}). Note that the constant density
$n$ of the plasma at equilibrium is equal to the density of dipole
moments. The polarization is automatically proportional to the gravitational
field. This is an attractive feature of the present formulation based on a
gravitational plasma; there is no need to invoke a mechanism to align $\Pi^i$
with $g^i$. Following the same reasoning as in Sec.~\ref{sec2}, the
polarization~(\ref{respolar}) gives rize to the density of polarization
masses~(\ref{rhopol}), which is added to the RHS of the Poisson law,
Eq.~(\ref{MONDpol}). The MOND equation~(\ref{MONDeq}) readily follows.

Let us comment on the field equation~(\ref{divf}) for the supposedly
fundamental internal interaction $f^i$. In the absence of gravity we have
$\chi=-1$, hence Eq.~(\ref{divf}) becomes that of a \textit{negative} Poisson
equation corresponding to a negative Newtonian force (with
``anti-gravitational'' constant $-G$). On the other hand, in strong gravity
($g/a_0\rightarrow\infty$) we have $\chi\rightarrow 0$ and we see that the
strength of the internal field $f^i$ becomes infinite. Therefore the plasma
gets locked in its undisturbed equilibrium state for which $\pi^i=0$ (strictly
speaking this is true if $g\chi\rightarrow 0$ when $g\rightarrow\infty$). In
this limit the dipolar medium is inactive; there is no induced polarization
and the Newtonian law holds.

Note that there is clearly some amount of fine tuning in the present model
(and the one of paper~II). Namely the function $\chi(g)$ is not determined
from first principles within the model but is tuned to agree with the
phenomenology of MOND. In particular $\chi(0)=-1$ is not ``explained'' in the
model but comes from astronomical observations. Similarly $\chi(\infty)=0$ is
imposed in order to recover the Newtonian regime where it is observed to be
valid.

To conclude, the phenomenology of MOND suggests the existence of a
polarization mechanism at work at the scale of galactic structures, and which
could be viewed as the gravitational analogue of the electric polarization of
a dielectric material. The dark matter would consist of the polarization
masses associated with some gravitational dipole moments aligned with the
gravitational field of ordinary masses. We find that the properties of the
dipolar dark matter are governed by the internal non-gravitational force
linking together the constituents of the dipolar medium. The formulation of
this medium in terms of an oscillating gravitational plasma polarized in the
gravitational field is particular attractive. In paper~II we shall show how
the notion of dipolar particle can be made compatible with the framework of
general relativity.

\ack It is a pleasure to thank C\'edric Deffayet, Gilles Esposito-Far\`ese,
Moti Milgrom and Jean-Philippe Uzan for interesting discussions.

\end{document}